\documentclass[12pt,epsf]{article}

\usepackage{amssymb,amsmath}

\newcommand{\bc}{\begin{center}}
\newcommand{\ec}{\end{center}}
\newcommand{\bd}{\begin{displaymath}}
\newcommand{\ed}{\end{displaymath}}
\newcommand{\be}{\begin{equation}}
\newcommand{\ee}{\end{equation}}
\newcommand{\ba}{\begin{array}}
\newcommand{\ea}{\end{array}}
\newcommand{\bea}{\begin{eqnarray}}
\newcommand{\eea}{\end{eqnarray}}
\newcommand{\bt}{\begin{tabular}}
\newcommand{\et}{\end{tabular}}

\newcommand{\bp}{\begin{picture}}
\newcommand{\ep}{\end{picture}}
\newcommand{\bfi}{\begin{figure}}
\newcommand{\efi}{\end{figure}}

\begin{document}

\title{\huge \bf {Production and Decay of 750 Gev state of 6 top and 6 antitop quarks}}

\author{ C. D. Froggatt ${}^{1}$ \footnote{\large\, colin.froggatt@glasgow.ac.uk} \ \ \
H.B.~Nielsen ${}^{2}$ \footnote{\large\, hbech@nbi.dk} \\[5mm]
\itshape{${}^{1}$ Glasgow University, Glasgow, Scotland}\\[0mm]
\itshape{${}^{2}$ The Niels Bohr Institute, Copenhagen, Denmark}}

\maketitle

\begin{abstract}
Crude estimates are made of the
branching ratios and pair production
cross-section  for our previously
proposed bound state S of six top and six
antitop quarks
identified with the
diphoton excess recently observed in
ATLAS and CMS.
We estimate the pair production cross
section to be approximately 12 times
that for  fourth family quarks.  Hence we
predict $\sigma(pp \rightarrow SS +
anything) \approx 2$ pb at 13 TeV and
an increase by a factor 10 over the
cross section at 8 TeV.
Crude estimates of the main branching
ratios relative to the diphoton decay
give $\Gamma(S\rightarrow \gamma+\gamma)
\propto 1$,
$\Gamma(S\rightarrow t + \bar{t})\propto 378$,
$\Gamma(S \rightarrow gluon+gluon)\propto 117$,
$\Gamma(S\rightarrow Higgs+Higgs)\propto 15$,
$\Gamma(S\rightarrow W+W)\propto 30$ and
$\Gamma(S\rightarrow Z+Z) = 15$.
 These estimates are
consistent
with the LHC bounds at 8 TeV within a
factor 1.25.
We expect the $S \rightarrow \gamma\gamma$ events
to be produced together with another S resonance
decaying typically into top-antitop or gluon-gluon jets.

\end{abstract}

\date{}


\newpage

\thispagestyle{empty}
\section{Introduction}
\label{introduction}
We shall here explore the possibility
that the diphoton excess in the inclusive
$\gamma\gamma$ spectrum,
recently found by the ATLAS and CMS
collaborations \cite{ATLASNOTE,CMS13},
with a mass of 750 GeV
can be a bound state of particles already
present in the Standard Model, namely
a bound state of 6 top + 6 antitop quarks. Thus we would need {\em no new
fundamental particles, interactions or
free parameters}
beyond the Standard Model to explain this peak, which otherwise looks like
``new physics''!

For several years we have worked on the
somewhat controversial idea \cite{1,3,10,
Kuchiev,Richard,Kuchiev2}
that the exchange of Higgses and gluons
between 6 top and 6 antitop quarks
provides sufficiently strong attraction
between these quarks for
a very light (compared to the mass of 12
top quarks) bound state S to be formed.
The 6 tops + 6 antitops are all
supposed to be in the 1s state in the
atomic physics notation and, because of
there being just 3 colors and
2 spin states for a top-quark, this is
the maximum number allowed in the 1s shell.

Further speculations around this bound
state were mostly built up under the
assumption of a hoped for new
 principle -- the multiple point
principle  \cite{MPP1,MPP2,MPP3} --
from which we actually predicted the
mass of the Higgs boson long before it
was found \cite{Higgsmass}. This principle says that
there shall be several
phases of space (i.e. several vacua) with
the same energy density. One of these
should have a condensate of the
bound states S. It was even speculated
then that such a condensate -- or new
vacuum -- could form
the interior
of balls, containing highly compressed
ordinary matter, which make
up the dark matter
\cite{PRLdark,Tunguska,Supernova}. Thus
the discovery, if confirmed, of the
bound state S could support a theory, in
which dark matter could be incorporated
into a pure Standard Model theory, only
adding the multiple point principle,
which predicts the values of coupling
constants but otherwise without new
physics.

It is the main goal of the present
article to crudely estimate the relative
decay rates
for the important channels and the pair
production rate of the particle S.
The experimental cross sections
$\sigma(p+p\rightarrow \gamma +\gamma
+anything) $ for the diphoton excess
at 750 GeV
are  $(6 \pm 3)$ fb at CMS and
$(10\pm 3)$ fb at ATLAS for $\sqrt{s} = 13$
TeV, while at 8 TeV they are respectively
$(0.5 \pm 0.6)$ fb and $(0,4 \pm 0,8)$ fb
\cite{ATLASNOTE,CMS13,CMS8,ATLAS8,
Franceschini}. For a singly produced
resonance by gluon fusion,
one expects a ratio $r=5$ for the
production cross section at 13 TeV
compared to that at 8 TeV. For pair
production,
like in our model, one rather expects a
ratio of $r=10$, which fits the data
better.

\section{Decay Diagrams}
\label{decay}

Of course whatever the outgoing particles
from the decay, they must
somehow couple to the (anti)top quarks
that are the constituents of the bound
state S. Also with the large number of
constituents it is needed that the
majority of these constituents simply
disappear in annihilations, in most cases
into nothing -- not even energy and momentum.

Let us consider the construction of a
diagram
for such  a decay, by first ignoring the
outgoing
particles. So, we imagine an annihilation
for each
top quark having some spin and color
with the
antitop quark having just the opposite
spin and color.
This can be represented by a blob
symbolizing the
bound state (Bethe Salpeter) wave
function, with
six loops attached to it representing
the 6 top
quarks annihilating with the 6 antitop
quarks.
A realistic decay diagram is now formed by
attaching an outgoing line to one of the
six loops for each of the final state
particles.
The important decay channels contain
{\em  two} final
state particles, which can be attached to
the
{\em same} loop or to {\em two different}
loops.
The top + antitop decay mode has to be
treated separately

However it is important to take
into account that
several of the decay products carry
global conserved quantum numbers which,
in the two loop case,
need to be transferred from one top
antitop annihilation loop to the next one.
For instance the Higgs carries
weak $SU(2)$ charge
(also weak hypercharge).
Only the photon, or rather its
and $Z^0$'s  component
coupling to the weak hypercharge, is the
exception and carries no global
conserved charge (apart from charge
conjugation).
In order to achieve such a transfer of
the global
charge from one loop attached to a final
state particle carrying the charge
to the loop attached to the other final
state particle, another particle
(e.g. a Higgs particle) has
to be exchanged between the loops.

We shall postpone the discussion of the
particles having a global conserved
charge until section \ref{twoloop} and
consider here the main factors, in the
diphoton decay, suppressing
the amplitude for both photons coming
via the same loop $T_1$ and via 2
different loops $T_2$ respectively.

\begin{itemize}
\item{{\bf  Suppression factor $T_1$ for
one loop giving the final state}}

In order that a pair of (anti)top
quarks -- with the compensating color and
spin of course -- can disappear into zero
four momentum and zero spin,
they should be sufficiently close, i.e.
at the same point, or at least within the
distance of a top quark Compton wavelength.
Thus the amplitude for such
total annihilation has to be proportional
to the amplitude for finding them
at the same point. We may get an idea of
this amplitude
by a dimensional argument considering the
wave function for the position of a
top quark
relative to its compensating antitop
quark $\Psi(\vec{r})$. In a Gaussian
wave function ansatz,
we can write such a wave function as the
product of one associated with each of the
three dimensions $ \Psi(\vec{r}) = \psi(x) \psi(y) \psi(z)$. For these wave
functions we need an estimate of the
radius $b/m_t$ of the bound state.
In our earlier work \cite{10} we
estimated $b \sim 1$ from the requirement
that the bound state
should bind to have a small mass compared
to the sum of the constituent quark masses.
This estimate was refined a bit further
in \cite{Larisanew} to give $b=2.3$.
The definition of the radius $b/m_t$
which we use is that
\begin{equation}
 <x^2> = \frac{<\vec{r}^2>}{3} =
\frac{b^2}{m_t^2}
\end{equation}
where this x-fluctuation should be that
of a single (anti)top relative to
the whole center of mass.
So the normalized wave function along
the x-axis for one constituent becomes
\begin{equation}
 \psi(x) = \sqrt{\frac{m_t}{\sqrt{2\pi}b}}\exp\left(-\frac{m_t^2}{4b^2}x^2\right).\label{wf}
\end{equation}
The relative wave function for two (say
compensating) constituents is given by
the convolution in x-space of two such
wave functions:
\begin{equation}
 \psi_{rel}(x) = \sqrt{\frac{m_t}{2\sqrt{\pi}b}}\exp\left(-\frac{m_t^2}{8b^2}x^2\right).
\end{equation}
The amplitude for finding the two (anti)top quarks with the same x coordinate is thus
$\psi_{rel}(0) = \sqrt{\frac{m_t}{2\sqrt{\pi}b}}$.
Assuming, on dimensional grounds, that
the top mass gives the relevant scale, we
extract the
typical suppression factor
\begin{equation}
 \xi_x = \psi_{rel} (0)/\sqrt{m_t}
= \sqrt{\frac{1}{2\sqrt{\pi}b}}.
\end{equation}
for the annihilation of the pair of
(anti)top quarks having this relative
wave function.
The amplitude for the decay of the whole
bound state gets suppressed by such a
factor for each totally annihilating
pair and for
each dimension of space. Thus the
amplitude of suppression from one
annihilating
pair becomes
\begin{equation}
 \xi = \xi_x\xi_y\xi_z =
\left (\sqrt{\frac{1}{2\sqrt{\pi}b}}\right )^3
=\left(4\pi b^2\right )^{-3/4}
= 0.043.
\end{equation}
The one loop decay has one more pair
annihilating into the vacuum
than the two loop decay and is thus
suppressed by an extra factor
of $\xi$.

When the various (anti)top pairs
annihilate, they
can deliver their full energy to one loop
or another from which the
final state particles arise. In the
{\em one}
loop case all the 6 pairs deliver
their energy to the surviving pair and
this has a combinatorial chance of
$(1/6)^6$. On the other hand in the case
of two loops,
each surviving pair having got
contributions of energy from 3 pairs, we
would have the probability
\begin{equation}
(1/6)^6 *\left ( \begin{tabular}{c} 6\\3 \end{tabular} \right )/2 = (1/6)^6*10.
\label{probability}
\end{equation}
Thus there is a combinatorial suppression
by a factor of 1/10 of the {\em one} loop decay
relative to the {\em two} loop decay.

Since the six (anti)top pairs in the
bound state are bosons, in states that
only differ by color and spin, the
interference terms between the decay
amplitudes arising from various
permutations of these pairs should get
the same phase. The
resulting constructive interference means
that the combinatorial factor, 1/10,
suppressing the one loop case relative to
the two loop case should apply to
the decay {\em amplitude}.

Collecting together the above two
suppression factors,
we get an overall suppression factor for
the {\em one} loop case of
\begin{equation}
 T_1 = \xi/10 = 0.0043.
\end{equation}

\item{{\bf Suppression factor $T_2$ for two  loops giving the final state}}

If both final state particles originate from the {\em same} (anti)top quark
annihilation pair the large spatial momentum of these particles is achieved by
the exchange of a fundamental particle (a top quark) in the loop between them,
while their collected three momentum is just zero. So, in the one loop
case, the bound state wave function causes no severe suppression of the emission
amplitude\footnote{The effects of a residual suppression could be incorporated by
reducing the value of $T_1$ and hence $\epsilon$ in eq.~\ref{f12} which, in any case,
has a large uncertainty.}.
If, however, the two final state particles arise from
two {\em different} (anti)top quark
annihilations, the final large spatial
momentum of the order of $m_S/2$ for each
of the decay particles has
to result from the momentum of the
antitop and top quarks annihilating
into that final state particle. This
means that, in momentum representation,
the emission amplitude is proportional to
the momentum wave function at the momentum being say
$\vec{p}=(m_S/2,0,0)$.
We must now estimate the width of the
momentum distribution for the
two (anti)top pairs, which emit the final
state particles, after the other 4 pairs
have annihilated into the vacuum. So we
imagine that three of the (anti)top
pairs give their momenta to one final pair, while the other three give their
momenta to the other final pair. This
would mean that the momentum distribution
of one of the final pairs, or
equivalently of one final state particle,
is given
by the convolution of the momentum
distributions for 3 pairs, meaning 6
(anti)top quarks. Fourier transforming
the single constituent
wave function (\ref{wf}), we obtain
\begin{equation}
 \tilde \psi (p_x) \propto
\exp\left(- \frac{p_x^2b^2}{m_t^2} \right).
\end{equation}
The distribution of the final state momentum, prior to its emission, is like that
of the sum of 6 single constituent momenta $\vec{p}_{fin} = \sum_{j=1}^6 \vec{p}_j $.
Thus we obtain the dependence of the final state particle wave function on
$p_{x \; fin}$:
\begin{equation}
\tilde \psi_{outgoing} (p_{x \; fin}) \propto
\exp\left(- \frac{p_{x \; fin}^2b^2}
{6m_t^2} \right)\label{wffin}.
\end{equation}
Substituting $p_{x \; fin} = m_S/2 $ into
(\ref{wffin}),
we obtain the wave function suppression
amplitude $T_2$ for the
case of the two final state particles (each with its own wave function) coming from two {\em different}
annihilating pairs to be:
\begin{equation}
T_2 = \left[\exp
\left(- \frac{(m_S/2 )^2b^2}{6m_t^2}
\right)\right]^2
 = \left[\exp(-4.1)\right]^2 = 0.00025 \label{T2}
\end{equation}

Now, however, the Gaussian wave function
form is {\em not trustable} for such a
large
momentum. Rather we should use a more
realistic form of the wave function like
the
one from say an atomic s-wave. The
Fourier transform of such a more realistic
wave function has the form
\begin{equation}
\tilde\psi(\vec{p}) \propto \left (\frac{1}{1 + \frac{1}{2} * \frac{(\vec{p} )^2b^2}{6m_t^2}}\right )^2.
\end{equation}
Inserting the momentum value
$|\vec{p}| = m_S/2$ for the decay
we obtain, for the atom-like wave
function form, the suppression factor
\begin{equation}
 T_2 =
 \left ( 1 + \frac{1}{2} *
\frac{(m_S/2 )^2b^2}{6m_t^2}\right )^{-4}
 = 0.106^2 = 0.011. \label{f11}
\end{equation}

\end{itemize}

Thus the ratio of the diphoton decay amplitude via one loop
to the photon pair decay amplitude (or better the component of the photon
coupling to the weak hypercharge) via two loops becomes:
\begin{equation}
 \epsilon= \frac{T_1}{T_2}
= \frac{0.0043}{0.011} = 0.39. \label{f12}
\end{equation}

To estimate the uncertainty in the ratio
$\epsilon$ we take
the parameter $\xi_x$, giving $\xi= \xi_x^3$, which is  estimated only by dimensional
arguments, to have an uncertainty of the
order of  a factor 2.
The counting of the combinatorics for
getting the energy collected on the one
or two pairs
we take to have an uncertainty by a factor
$10/3 = 3.3$ in the amplitude (allowing $\left( \begin{tabular}{c} 6\\3 \end{tabular} \right)$
to be replaced by $\left ( \begin{tabular}{c} 4\\2 \end{tabular} \right)$ in eq.~\ref{probability}).
The wave function(s) we take to be
uncertain by a factor 2 in the exponent,
meaning by a factor $\exp(4.1/2)$.
Adding the squares of the logarithms of
these uncertainties leads to
the total uncertainty for our estimate of
the ratio $\epsilon$

\begin{equation}
 \Delta(\ln \epsilon) = \sqrt{ (\ln 8)^2 + (\ln 3.3)^2 + (4.1/2)^2} = \sqrt{4.3 +1.4 +4.2} = \sqrt{9.9} = 3.1.
\end{equation}
This means that that this ratio $\epsilon$ is uncertain by a factor $\exp(3.1) = 20 $.

\section{Branching Ratios}

We shall now estimate the decay rates of the
bound state S into the important 2 body channels
relative to the diphoton decay rate.
Firstly we consider the one loop decay mechanism,
which will be dominant for large values of
$\epsilon=T_1/T_2$ and then the two loop decay mechanism
which will be dominant for small values of $\epsilon$.

\subsection{One Loop Case} \label{oneloop}

The main decay channels, apart from the
$S \rightarrow t\bar{t}$ decay, correspond
to the same diagram structure differing only in
the couplings of the final state particles to the
annihilating top antitop loop. So their relative
decay rates should essentially be given by the ratios
of the coupling strengths involved. We shall now
introduce effective couplings for the various final state
particles, analogous to the fine structure constant
$\alpha=e^2/4\pi$ in QED. These effective
couplings are used in Table \ref{tableonepair}
to roughly estimate the relative decay rates of the bound
state S, via the one loop mechanism, into the dominant
2 body final states.

\begin{table}
\begin{tabular}{|c|c|c|c|c|}
\hline
Final state $f$ & Bound & Relative prediction& $\frac{\Gamma(S\rightarrow f)}
{\Gamma(S\rightarrow \gamma\gamma)}$& Comment\\
\hline
$\gamma\gamma$ & $<0.8(r/5)$ & $(4\alpha/9)^2 = 1.2*10^{-5}$ & 1&\\
gluon + gluon & $<1300 (r/5)$& $8(\alpha_s/6)^2=2.3*10^{-3}$ & 190&\\
Higgs + Higgs &  $<20(r/5)$ & $ \alpha_h^2/4 =3*10^{-4}$ & 25&  Higgs-particles\\
ZZ&$<6(r/5)$& $\alpha_h^2/4 =3*10^{-4}$ & 25& longitudinal\\
WW & $<20(r/5)$& $\alpha_h^2/2 = 6*10^{-4} $& 50&longitudinal\\
$Z\gamma$ & $< 2(r/5)$ & $2(4\alpha/9)^2*0.92=2.2*10^{-5} $&1.8 &\\
ZZ& $<6(r/5)$ & $(4\alpha/9)^2*(0.92)^2= 1.0*10^{-5}$&0.8&transverse\\
WW & $< 20(r/5)$ &$2(0.54\alpha)^2 =3.5*10^{-5}  $ &3& transverse \\
top + antitop & $< 300 (r/5)$ & $3\alpha_{t\bar{t}}^2=5.9*10^{-3}$ & 494&\\
\hline
\multicolumn{3}{| r}{ $\Gamma_{total}(S)/\Gamma(S\rightarrow\gamma\gamma)$:} & 791 &\\
\hline
\end{tabular}
 \caption{Assuming dominance of {\em one} top antitop pair giving the final state,
   relative predictions are given for the
partial decay widths of S and for
the branching ratios relative to the
diphoton decay width compared to the
experimental upper bounds from
ref. \cite{Franceschini}. In our model $r \simeq 10$.}
\label{tableonepair}
\end{table}

\begin{description}

\item[Photon and transverse Z.]
The electric charge of the top quark is $q = 2e/3$
and the effective coupling for of the photon to the
$t\bar{t}$ loop is $4\alpha/9$.

The corresponding effective coupling of Z to the
$t\bar{t}$ loop is
\begin{equation}
\frac{\alpha}{2\sin^2\theta_W\cos^2\theta_W}
\left[\left(\frac{1}{2} - \frac{2}{3}\sin^2\theta_W\right)^2
+ \left(-\frac{2}{3}\sin^2\theta_W\right)^2\right]
= \frac{4\alpha}{9}*0.92.
\end{equation}
We take $\alpha = 1/129$ and the Weinberg angle to be
given by $\sin^2\theta_W = 0.23$.

\item[Gluon.]
The vertex for a gluon of color $i$ coupling to a top quark is
$g_s\lambda^i/2$. Averaging over the colors of the top quark,
the effective coupling of the gluon to the $t\bar{t}$ loop becomes
\begin{equation}
\frac{\alpha_s}{3}Tr\left(\frac{\lambda^i}{2}\right)^2
= \frac{\alpha_s}{6}.
\end{equation}
We take $\alpha_s = 0.1$ and then sum over the 8 color states
of the gluon.

\item[Higgs and longitudinal $W^{\pm}$ and $Z^0$.]
According to the Goldstone Boson Equivalence Theorem \cite{Chanowitz},
in the high energy limit the couplings of the longitudinal
$W^{\pm}$ and $Z^0$ become equal to those of the corresponding
eaten Higgs fields. The Higgs field coupling to the $t\bar{t}$
loop is
\begin{equation}
\alpha_h = \frac{g_t^2/2}{4\pi} = 0.035
\end{equation}
where $g_t$ is the top quark Yukawa coupling constant. Also
the Higgs field has only one helicity state unlike the transverse
gauge bosons and the top quark.

\item[Transverse $W^{\pm}$.]
The $W^{\pm}$ gauge fields are formed from two real fields,
$W_1$ and $W_2$, lying in the adjoint representation of SU(2).
So their effective coupling to the $t\bar{t}$ loop is
\begin{equation}
\frac{1}{2}*\frac{\alpha}{\sin^2\theta_W}\left(\left(\frac{\sigma^i}{2}\right)^2\right)_{t_Lt_L}
= \frac{\alpha}{8\sin^2\theta_W} = 0.54\alpha,
\end{equation}
where the extra factor of 1/2 is due to $W^{\pm}$ only
interacting with left-handed top quarks. The final sum
over $i = 1,2$ gives a factor of 2 in the decay rate.

\item[Top antitop.]
In principle one top and one antitop  quark can simply escape,
while the other 5 pairs annihilate into the vacuum. However
this would happen with an amplitude suppressed by the square
of the wave function for finding each of the escaping quarks having
spatial momentum $m_S/2$, i.e. suppressed by the factor $T_2$ of
eq. \ref{f11}. If the top and antitop exchange a gluon between
themselves, as they escape, they can get most of the kinematically
needed spatial momentum exchanged through that gluon propagator
and thus the emission could occur at first from where the wave
function is not so suppressed. We assume that, crudely, having
such an exchange of some fundamental particle essentially eliminates the
suppression due to the wave function, as we do for the other one loop
processes (see footnote 1). Thus in amplitude we expect that, in
compensation for the $\alpha_s$ factor due to the gluon exchange,
the wave function suppression factor $T_2$ can be dropped.
So, for the top antitop decay we consider a one loop diagram
consisting of a gluon exchange between the quarks. We therefore
assume that the ratio of the top antitop decay rate to the
other one $t\bar{t}$ loop decays is well estimated by taking the ratio
of the coupling strengths involved. The gluon exchange diagram
involves the quadratic Casimir of the quark representation (rather
than the index involved in the $S\rightarrow gg$ decay) and
the effective coupling for the production of a quark of color $a$ becomes
 \begin{equation}
\alpha_{t\bar{t}} = \frac{\alpha_s}{3}\sum_i\left(\left(\frac{\lambda^i}{2}\right)^2\right)_{aa}
=\frac{4\alpha_s}{9}.
\end{equation}
Finally we sum over the 3 color states of the quark.
\end{description}

\subsection{Two Loop case} \label{twoloop}

We now consider the case where the two final state particles are
emitted by two different quark pairs. The gluon-gluon decay channel
is special, because the gluons can be emitted from two ``crossed'' loops,
in which the two (anti)top quark pairs swop color indices.
For the other channels (apart from $S \rightarrow t\bar{t}$
decay) the two (anti)top pairs separately annihilate into a colorless final
state particle. However, for particles with a global conserved quantum number
(i.e. all the other relevant final state particles except for the hypercharge
coupling part of the photon or of the $Z^0$) the emission from
two different pairs is forbidden, unless a particle carrying the global
quantum number is exchanged between the two pairs. This means that a
diagram for e.g. $S \rightarrow Higgs + Higgs$ will, in addition to the $6-2 = 4$
pairs annihilating into the vacuum, have two (anti)top pairs coming out of the
bound state vertex, exchanging a Higgs particle and producing the 2 final
state Higgs particles.

The diagrams for the emission of colorless final state particles with global
charges all look very similar and even similar to the diagram
for $S \rightarrow t + \bar{t}$ decay, in which an inner top quark loop
exchanges a gluon with each of the outgoing $t$ and $\bar{t}$ quarks.
They namely all have three loops in addition to the four (anti)top loops
annihilating into nothing. Thus we assume that these various diagrams give decay
rates essentially in the ratios of the coupling strengths involved.
We note that the wave function suppression factor $T_2$ is partially relaxed by
the additional particle exchanged between the two loops. However the emission
vertices for the two final state particles are connected by 3 propagators in series
rather than by only one propagator as in the one loop case. The transfer of the
final state momentum through these 3 propagators between the vertices is less
effective and thus the relevant momenta for the bound state wave function
are larger than in the one loop case. We assume that, crudely, the wave function
suppression factor $T_2$ is thereby relaxed to a factor of $\sqrt{T_2}$
and not replaced by unity as was done in the one loop case. On the other hand, 
the diphoton and gluon-gluon decay amplitudes retain the full wave function 
suppression factor $T_2$.

The diagrams for the emission of the weak hypercharge coupled components
of the photon and $Z^0$ and the ``crossed" loop diagram for gluon decay
are a bit simpler, with only two loops in addition to the 4 annihilation
into nothing loops. So we have to adopt a rule to compare the two loop
diagrams with the three loop diagrams. We shall assume that the addition
of a loop to a diagram roughly leads to an extra fine structure constant factor
$\alpha_x$ appropriate to the exchanged particle times the old diagram.
We could of course have used another expansion parameter instead of $\alpha_x$;
for example $\alpha_x/\pi$ which would have led to a reduction in the
Higgs + Higgs, ZZ, WW and top + antitop decay branching ratios in
Table \ref{tabletwopair} by an order of magnitude. The gluon + gluon
branching ratio does not suffer from this uncertainty.

Let us now consider the effective couplings to be used in Table \ref{tabletwopair}
to roughly estimate the relative decay rates of the bound
state S, via the two loop mechanism, into the dominant
two body final states.

\begin{table}
\begin{tabular}{|c|c|c|c|c|c|}
\hline
Final state $f$ & Bound & Relative prediction& $\frac{\Gamma(S\rightarrow f)}
{\Gamma(S\rightarrow \gamma\gamma)}$& Comment\\
\hline
$\gamma\gamma$ & $<0.8(r/5)$ & $(0.236\alpha)^2 = 3.35*10^{-6}$ & 1 &\\
gluon + gluon & $<1300 (r/5)$& $8(\alpha_s/18)^2=2.5*10^{-4}$ & 74 &\\
Higgs + Higgs &  $<20(r/5)$ & $\alpha_h^4/(4T_2) = 3.4*10^{-5}$ & 10 &  Higgs-particles\\
ZZ&$<6(r/5)$& $\alpha_h^4/(4T_2) = 3.4*10^{-5}$ & 10 & longitudinal\\
WW & $<20(r/5)$& $\alpha_h^4/(2T_2) = 6.8*10^{-5}$&  20 &longitudinal\\
$Z\gamma$ & $< 2(r/5)$ & $2(0.236\alpha)^2\tan^2\theta_W=2.0*10^{-6}$ & 0.6 &\\
ZZ& $<6(r/5)$ & $(0.236\alpha)^2\tan^2\theta_W = 3.0*10^{-7}$&0.09 &transverse\\
WW & $< 20(r/5)$ &$2(0.54\alpha)^4/T_2 = 6*10^{-8}  $ &  0.02 & transverse \\
top + antitop & $< 300 (r/5)$ & $3\alpha_{t\bar{t}}^4/T_2=1.06*10^{-3}$ & 316&\\
\hline
\multicolumn{3}{| r}{ $\Gamma_{total}(S)/\Gamma(S\rightarrow\gamma\gamma)$:} & 432&\\
\hline
\end{tabular}
 \caption{Assuming dominance of
{\em two} top antitop pairs giving the
final state,
   relative predictions are given for the
partial decay widths of S and for
the branching ratios relative to the
diphoton decay width compared to the
experimental upper bounds from
ref. \cite{Franceschini}. In our model $r \simeq 10$.}
\label{tabletwopair}
\end{table}

\begin{description}

\item[Photon and transverse Z.]
The hypercharge coupled superposition of the photon and $Z^0$
is described by the field $B_{\mu} = \cos\theta_WA_{\mu} - \sin\theta_WZ_{\mu}$.
It couples with an average squared charge $[(2/3)^2 + (1/6)^2]/2 = 0.236$
to a top quark. The two loop diphoton decay is dominated by the production
of this $B_{\mu}$ component and so the effective coupling for the photon is $0.236\alpha$.

The corresponding effective coupling of Z is $0.236\alpha\tan\theta_W$.

\item[Gluon.]
Averaging over the colors of the two (anti)top pairs, the effective
coupling of a gluon of color $i$ for the ``crossed" diagram is
\begin{equation}
\frac{\alpha_s}{9}Tr\left(\frac{\lambda^i}{2}\right)^2
= \frac{\alpha_s}{18}.
\end{equation}

\item[Higgs, longitudinal $Z^0$, $W^{\pm}$ and top antitop.]
We use the same effective couplings as in the one loop case.

\end{description}

\subsection{Benchmark Model}

We estimated the ratio of  the photon
pair decay amplitude via one loop to the hypercharged component of
the photon pair decay amplitude via two loops
in eq.~(\ref{f12}) to be $\epsilon =0.39$, but with an uncertainty
of order 20. So it is not clear which mechanism dominates.
We take, as
a benchmark model, the combination of the two loop decay rates from
Table \ref{tabletwopair} plus $\epsilon^2 =0.15$ times the one loop decay
rates from Table \ref{tableonepair}, neglecting possible interference terms
between the two decay mechanisms, and calculate the corresponding relative
decay branching ratios given in Table \ref{tablebenchmark}.

\begin{table}
\begin{tabular}{|c|c|c|c|}
\hline
Final state $f$ & Bound
& $\frac{\Gamma(S\rightarrow f)}
{\Gamma(S\rightarrow \gamma\gamma)}$& Comment\\
\hline
$\gamma\gamma$ & $<0.8(r/5)$
& 1&\\
gluon + gluon & $<1300 (r/5)$
& $ 117$&\\
Higgs + Higgs &  $<20(r/5)$
& $15$&  Higgs-particles\\
ZZ&$<6(r/5)$
& $15$& longitudinal\\
WW & $<20(r/5)$
& $30$&longitudinal\\
$Z\gamma$ & $< 2(r/5)$
&1.0 &\\
ZZ& $<6(r/5)$
&0.3&transverse\\
WW & $< 20(r/5)$
&$1.1$& transverse \\
top + anti top & $< 300 (r/5)$
& $378$&\\
\hline
\multicolumn{2}{| r}{ $\Gamma_{total}(S)/\Gamma(S\rightarrow\gamma\gamma)$:} &558&\\
\hline
\end{tabular}
 \caption{Benchmark model with
$\epsilon^2= 0.15$.
   Predictions are given for the
decay branching ratios of S relative to the
diphoton decay width and compared to the
experimental upper bounds from
ref. \cite{Franceschini}. In our model $r \simeq 10$.}
\label{tablebenchmark}
\end{table}

\section{The Production Cross
Section}
\label{production}

Our bound state particle S would be
produced in pairs by the
gluon fusion process $gg \rightarrow SS$.
(We think that producing just one S
would be very suppressed by essentially
our $\xi$ factors.)
We shall estimate the
cross section for this process by using
an impulse approximation,
according to which the two gluons
interact with just one quark
inside the bound state.

Let us first consider the scattering
process related to  $gg \rightarrow SS$
by crossing symmetry: the scattering of a
gluon by the bound state S in the
impulse approximation \cite{belinfante}.
That is to
say we take it, that a gluon scatters on
just one of the constituents.
The scattering amplitude is then written
as a matrix element
of the constituent density in the bound
state S multiplied by the amplitude for
the gluon
scattering on one constituent. The matrix
element of the density
of constituents, which we call the form
factor $F(p^2)$, is the Fourier
transform of the spatial density of
constituents in S. Since there are twelve
constituent
(anti)top quarks we have $F(0) = 12$.
Here the four momentum $p$ -- which in
the  $gS \rightarrow gS$ process is
purely space-like -- is that delivered
from the gluon to the
S. Next we invoke crossing symmetry and
analyticity to continue the
amplitude for this $gS \rightarrow gS$
process into the process
of two gluons pair creating two S's:
$gg \rightarrow SS$.
The important four momentum $p$ for the
form factor $F(p^2)$ thereby
gets continued from its spatial value for
the $gS \rightarrow gS$ scattering to a time-like
value for the $gg \rightarrow SS$
pair production process.  In the
space-like region for
$p$ the form factor $F(p^2)$ will
typically fall off as $p^2$ becomes
numerically
bigger, since the density is typically
somewhat smooth and its
Fourier transform falls for large
(space-like) four momenta.
A naive analytical continuation therefore
suggests that going into
the opposite direction in the variable
$p^2$ -- namely the time-like
direction -- the form factor would grow
with larger and larger numerical
time-like $p$ until it meets a
singularity. This is what one sees
in cases like annihilation of say
$e^++e^-$ into two pions \cite{baldini} or into
proton antiproton \cite{Pacetti}. In the
case of pions the form factor $F(p^2)$
actually grows with increasing $p^2$ even
above the threshold
for producing the two pions; it first
meets its effective
singularity at the $\rho$-resonance,
after which the form factor turns
down for further increase in $p^2$. For
proton antiproton
the singularities are met already below
the threshold, so that above the threshold
the form factor is already decreasing
with numerically raising $p^2$.
The S states have rather strong Higgs fields
around them and hence we expect significant
(possibly resonant) dispersive contributions to
the S form factor from the SS channel above
threshold. So we expect the
form factor for S in the region above the
threshold
for pair production to be within a factor
of 10 or so from
its value $F(0)=12$ at $p^2=0$.
However we have to take into account that
the twelve different quarks and antiquarks
have rather different colors. So certain
combinations of the gluon color states
will not interact with certain quarks and
some couplings will get alternating signs.
It follows that the interference terms
between amplitude contributions from the
gluons
attaching to different (anti)quarks have
complicated signs. Hence we have decided
to ignore
the interference terms. So the square of
the sum of the 12 amplitudes from
each of the constituent quarks
gets approximated by the sum of the
numerical squares of the 12 terms
individually.

The actual estimation of the final cross section is
done by making use of the already calculated production cross
section for a fourth family (vectorlike) quark \cite{Karabacak}.
Thus we predict the production cross section of our bound state pair
should be $12$ times the pair production cross section for a
fourth family quark having
a mass of 750 GeV, but with an order of magnitude uncertainty.
The mass of 750 GeV is needed, rather than the top quark mass,
in order to give the correct kinematics
for the $gg \rightarrow SS$ process. Using the quark
pair production cross sections calculated in \cite{Karabacak}, we predict the cross section
$\sigma(pp \rightarrow SS + anything)$ to be $12*0.02 \sim 0.2$ pb at
$\sqrt{s} = 8$ TeV and $12*0.2 \sim 2$ pb
at $\sqrt{s}=14$ TeV.
The total decay width relative to the diphoton decay width
$\Gamma_{total}(S)/\Gamma(S \rightarrow \gamma\gamma)$ differs
by less than a factor of 2 between the one loop and two loop mechanisms.
Taking the benchmark value of
$\Gamma_{total}(S)/\Gamma(S \rightarrow \gamma\gamma) = 558$,
we predict the production cross section for S decaying to diphotons
at $\sqrt{s} = 13$ TeV to be 2/558 pb $\simeq 4$ fb, consistent with the
cross sections observed by ATLAS and CMS.

\section{Conclusion}

We have argued that the recently observed excess of diphoton events
with a mass of 750 GeV can be identified with the decay of our
proposed bound state S of 6 top quarks and 6 antitop quarks \cite{1,3,10},
without violating any experimental bounds for alternative decays.
In our model the resonance S is produced in pairs by two colliding gluons.
We estimate the pair production cross section, using the impulse approximation,
to be roughly a factor of 12 -- the number of (anti)top constituents -- times
bigger than the production cross section for a pair of fourth family
(vectorlike) quarks. This leads to an S production cross section of 2 pb
at 13 TeV. Using the benchmark model of Table 3 for the decay branching
ratios, a diphoton production cross section of order 4 fb is obtained in
agreement with ATLAS and CMS. We also expect the diphoton excess events
to be accompanied by another S resonance decaying typically into
top + antitop or gluon + gluon jets. This of course applies to the other
decay modes as well as the diphoton decay.

We have considered two decay mechanisms: one loop decay in which one
annihilating top antitop loop emits both final state particles and
two loop decay in which the two final state particles come from
different top antitop annihilating pairs. The decay by a simple
escape of a top and an antitop, each carrying spatial momentum $m_S/2$,
is suppressed by the bound state wave function. This suppression
is ameliorated by gluon exchange and the top + antitop
decay mode has the largest branching ratio in both cases.
The one loop decay rate relative to the two loop decay rate is
determined by the parameter $\epsilon^2$, which we estimated for our
benchmark model to be 0.15 but with a huge uncertainty factor of
order 400.

We note that the Higgs + Higgs and the longitudinal WW and ZZ decay
modes are close to their experimental bounds for both the one loop
and two loop decay mechanisms. It should also be remarked that the
total decay amplitude for our bound state S is suppressed by 4 or 5
factors of $\xi \sim 0.04$ (one for each (anti)top pair annihilating
into nothing) and thus a decay rate suppressed by a factor of $\xi^8$
or $\xi^{10}$. Hence, in our model, the resonance S would be very narrow
with a width much less than the energy resolution at LHC.

It must be admitted that our estimates of the branching ratios and cross section
are indeed very crude and subject to large uncertainties, but it is quite difficult
to do much better. Nevertheless it is in principle possible in our scheme to
calculate everything concerning the new particle, because our model is fundamentally
nothing but the Standard Model!

\section*{Acknowledgements}

HBN would like to thank Shi-Yuan Li of Shandong University for
discussions in 2008 of low multiplicity jet decays for
the 6 top + 6 antitop bound state; he also wishes to thank the
Niels Bohr Institute for his status as emeritus professor.
CDF would like to thank David Sutherland for helpful discussions;
he also wishes to acknowledge hospitality and support from Glasgow
University and the Niels Bohr Institute.


\begin{thebibliography}{99}

\bibitem{ATLASNOTE}
ATLAS Collaboration, ATLAS-CONF-2015-081,\\
https://atlas.web.cern.ch/Atlas/GROUPS/PHYSICS/CONFNOTES/ATLAS-CONF-2015-081/.
\bibitem{CMS13}
CMS Collaboration, CMS PAS EXO-15-004,\\
https://cds.cern.ch/record/2114808/files/EXO-15-004-pas.pdf.
\bibitem{1}
C.D.~Froggatt and H.B.~Nielsen,  Surveys High Energy Phys.
{\bf 18},  55-75 (2003)
[arXiv: hep-ph/0308144].
\bibitem{3}
C.D.~Froggatt, H.B.~Nielsen and L.V.~Laperashvili, Int. J. Mod.
Phys. A {\bf 20}, 1268 (2005) [arXiv: hep-ph/0406110].
\bibitem{10}
C.D.~Froggatt and H.B.~Nielsen, Phys. Rev. D {\bf 80}, 034033
(2009) [arXiv:0811.2089].
\bibitem{Kuchiev}
M.~Yu.~Kuchiev, V.~V.~Flambaum and E.~Shuryak, Phys. Rev. {\bf D78} (2008) 077502 [arXiv:0808.3632].
\bibitem{Richard}
J.-M.~Richard, Few Body Syst. {\bf 45} (2009) 65 [arXiv:0811.2711].
\bibitem{Kuchiev2}
M.~Yu.~Kuchiev, Phys. Rev. {\bf D82} (2010) 127701 [arXiv:1009.2012].
\bibitem{MPP1}
D.~L.~Bennett, C.~D.~Froggatt and H.~B.~Nielsen, {\it Proc.~of the 27th
International Conference on High Energy Physics}, edited by P.~Bussey and
I.~Knowles (IOP Publishing Ltd, london, 1995), p.~557 [NBI-HE-94-44, C94-07-20].
\bibitem{MPP2}
D.~L.~Bennett, C.~D.~Froggatt and H.~B.~Nielsen,
{\it Perspectives in Particle Physics `94}, edited by D.~Klabu\u{c}ar, I.~Picek
and D.~Tadi\`{c} (World Scientific, Singapore, 1995), p.~255 [arXiv:hep-ph/9504294].
\bibitem{MPP3}
D.~L.~Bennett and H.~B.~Nielsen, Int. J. Mod. Phys. {\bf A9} (1994) 5155.
\bibitem{Higgsmass}
C.~D.~Froggatt and H.~B.~Nielsen
Phys. Lett. {\bf B368} (1996) 96 [arXiv.hep-ph/9511371].
\bibitem{PRLdark}
C.~D.~Froggatt and H.~B.~Nielsen, Phys. Rev. Lett. {\bf 95} (2005) 23130
[arXiv:asto-ph/051245].
\bibitem{Tunguska}
C.~D.~Froggatt and H.~B.~Nielsen, Int. J. Mod. Phys. {\bf A30} (2015) 1550066
[arXiv:1403.7177].
\bibitem{Supernova}
C.~D.~Froggatt and H.~B.~Nielsen, Mod. Phys. Lett. {\bf A30} (2015) 1550196
[arXiv:1503.01089].
\bibitem{CMS8}
CMS Collaboration, CMS-PAS-EXO-12-045.
\bibitem{ATLAS8}
ATLAS Collaboration,
Phys. Rev. {\bf D92} (2015) 032004 [arXiv:1504.05511].
\bibitem{Franceschini}
R.~Franceschini, G.~F.~Giudice, J.~J.~Kamenik, M.~McCullough, A.~Pomaral,
R.~Rattazzi, M.~Redi, F.~Riva, A.~Strumia and R.~Torre,
arXiv:1512.04933 [hep-ph].
\bibitem{Larisanew}
L.~V.~Laperashvili, H.~B.~Nielsen and C.~R.~Das,
 Int. J. Mod. Phys. A31
(2016) 1650029.
arXiv:1601.03231 [hep-ph].
\bibitem{Chanowitz}
M.~S.~Chanowitz and M.~K.Gaillard,
Nucl. Phys. {\bf B261} (1985) 379.
\bibitem{belinfante}
J.~G.~Belinfante,
Phys. Rev. {\bf 128} (1962) 2403
\bibitem{baldini}
R.~Baldini, S.~Dubnicka, P.~Gauzzi, S.~Pacetti, E.~Pasqualucci and Y.~Srivastava,
Nucl. Phys. {\bf A666} (2000) 38c.
\bibitem{Pacetti}
S.~Pacetti, R~B.~Ferroli and E.~Tomasi-Gustafsson,
Phys. Rept. {\bf 550-551} (2015) 1.
\bibitem{Karabacak}
D.~Karabacak, S.~Nandi and S.~K.~Rai,
Phys. Lett. {\bf B737} (2014) 341 [arXiv:1405.0476].

\end{thebibliography}
\end{document}